\documentclass[preprint]{aastex}
\usepackage{graphicx,amsmath}
\newcommand\msol{\ensuremath{\,\mbox{\it M}_{\odot}}}

\newcommand\lta{\mathrel{\hbox{\raise 0.6 ex \hbox{$<$}\kern
                   -1.8 ex\lower .5 ex\hbox{$\sim$}}}}
\newcommand\gta{\mathrel{\hbox{\raise 0.6 ex \hbox{$>$}\kern
                   -1.7 ex\lower .5 ex\hbox{$\sim$}}}}
 
\newcommand{\scrbox}[1]{\ensuremath{{\mbox{\scriptsize #1}}}}

\newcommand{\teff}{{\ensuremath{T_{\scrbox{eff}}}}}
\newcommand{\Msol}{\ensuremath{\,\mbox{\it M}_{\odot}}}
\newcommand{\Mstar}{\ensuremath{{\it \,M_{*}}}}

\newcommand{\Dturb}{\ensuremath{D_{\scrbox{T}}}}
\newcommand{\Kelvin}{\,\mbox{K}}
\newcommand{\MS}{main--sequence}
\newcommand{\gr}{\ensuremath{g_{\scrbox{rad}}}}
\newcommand{\tbcz}{{\ensuremath{T_{\scrbox{bcz}}}}}
\newcommand{\mbcz}{{\ensuremath{M_{\scrbox{bcz}}}}}

\renewcommand{\H}{\mbox{H}}
\newcommand{\He}{\mbox{He}}

\newcommand{\Fe}{\mbox{Fe}}

\newcommand{\Li}{\mbox{Li}}

\newcommand{\T}{_{\rm{T}}}

\shortauthors{Richard et al.}
\shorttitle{WMAPS and Li abundance}
\slugcomment{The Astrophysical Journal, accepted}
\begin{document}

\title{Implications of WMAP observations on Li abundance and stellar evolution models}

\author{O.~Richard\altaffilmark{1}, G.~Michaud and J.~Richer}
\affil{D\'epartement de Physique, Universit\'e de Montr\'eal,
       Montr\'eal, PQ, H3C~3J7, CANADA}
\email{Olivier.Richard@graal.univ-montp2.fr, michaudg@astro.umontreal.ca,
       jacques.richer@umontreal.ca}

\altaffiltext{1}{GRAAL UMR5024, Universit\'e Montpellier II,
                 CC072, Place E. Bataillon,
                 34095 Montpellier Cedex,
                 France}
%% If you wish, you may supply running head information, although
%% this information may be modified by the editorial offices.
%% The left head contains a list of authors,
%% usually a maximum of three (otherwise use et al.).  The right
%% head is a modified title of up to roughly 44 characters.
%% Running heads will not print in the manuscript style.




\begin{abstract}
The WMAP determination of the baryon$-$to$-$photon ratio implies, through Big Bang nucleosynthesis, a cosmological
Li abundance larger, by a factor of 2 to 3, than the Li abundance plateau observed in the oldest Pop II stars. 
It is however  inescapable that there be a reduction by a factor of at least 1.6 to 2.0 of the surface Li abundance
during the evolution of  Pop II field stars with $[\Fe/\H] \le -1.5$. 
That the observed Li be lower than cosmologically produced Li  
 is expected from stellar evolution models.  Since at turnoff most of the Li
abundance reduction is caused by gravitational settling, the presence of $^6\Li$ in some turnoff stars is also understood.
Given that the WMAP implications for Li cosmological abundance and the Li Spite plateau can be naturally 
explained by gravitational settling in the presence of weak turbulence, there appears little need for exotic physics 
as suggested by some authors.
Instead, there is a need for a better understanding of turbulent transport in the radiative zones of stars.
This requires simulations  from first principles.
Rather strict upper limits to turbulent transport are determined for the Sun and Pop II stars.

\end{abstract}

%\today

%% subject header.  Objects should be in the appropriate "individual"
%% headers (e.g. quasars: individual, stars: individual, etc.) with the
%% additional provision that the total number of headers, including each
%% individual object, not exceed six.  The \objectname{} macro, and its
%% alias \object{}, is used to mark each object.  The macro takes the object
%% name as its primary argument.  This name will appear in the paper
%% and serve as the link's anchor in the electronic edition if the name
%% is recognized by the data centers.  The macro also takes an optional
%% argument in parentheses in cases where the data center identification
%% differs from what is to be printed in the paper.


%% and \citet commands to identify citations. 

\section{Astrophysical context}
\label{sec:context}
Most low metallicity ($[Z/\H]<-1.5$) halo field stars with $6300 > \teff{} > 5500$ K have
nearly the same surface Li abundance (\citealt{SpiteSp82}).  This result has frequently been confirmed since
and has lead many to suggest that those stars have 
the cosmological Li abundance.

In stellar evolution models however, one must include all physical processes that follow from first principles, 
including gravitational settling which  leads to
a reduction in the surface Li abundance; deeper in, nuclear reactions destroy Li. 
This combination, together with the similarity of the Li concentration in stars of different \teff{} and $[Z/\H]$, 
led  \citet{MichaudFoBe84} to conclude that ``the measured lithium abundance is not 
exactly the original abundance.  It must have been reduced 
by a factor of at least 1.5 and more probably 2, from the abundance with 
which the star formed.'' (p. 212).  

On the other hand, one may argue that, with the measurements of WMAP, the cosmic microwave background determines 
the baryon$-$to$-$photon$-$ratio, 
$\eta$, \citep{CyburtFiOl2002} so that Big Bang models 
predict the original Li abundance \citep{CyburtFiOl2003}.  This predicted original Li abundance is a factor of 2$-$3 larger
than the Li abundance in stars of the plateau
 \citep{SpiteMaSp84}.  More precisely, the WMAP best fit, which includes the use of other data sets (see sec. 3 of 
\citealt{CyburtFiOl2003}) leads to 
\begin{equation}
  \label{eq:wmapli}
  N(^7\Li)/N(\H)= 3.82^{+0.73}_{-0.60} \times 10^{-10}.
\end{equation}

In this paper we assume the cosmological Li abundance to be given by equation (\ref{eq:wmapli}) and determine
the implications for stellar evolution models 
of Pop II stars.  To this end we first briefly review some pertinent Li observations (Sec. \ref{sec:observations}) and
 then (Sec. \ref{sec:evolution})
the surface Li abundance to be expected from a comprehensive series of stellar evolution models that treat particle transport in great detail 
\citep{RichardMiRietal2002,RichardMiRi2002,VandenBergRiMietal2002}.  In section~\ref{sec:discussion}, a comparison to observations
leads to a discussion  of the implications for turbulence modeling, of the
link to solar type stars as well a discussion of other points of view.  
In the conclusion section, by putting all results together, we emphasize that atomic diffusion may be the main cause 
of the reduction of \Li{} abundance between the original value given by WMAP and that observed in Pop II stars, 
with  turbulent transport acting as a perturbation.
The constancy of the \Li{} abundance as a function of \teff{} is then less surprising than originally thought.  
Potential observational tests are also suggested.

\section{$^6\Li$ and $^7\Li$ Observations}
\label{sec:observations}

In the  halo, \citet{SpiteSp82} have obtained  very similar Li abundances for  nearly all low metallicity dwarf stars with 
$\teff \geq 5500$ \Kelvin{}. 
Following those  original observations, the existence of the Li abundance plateau in 
Pop II halo stars has been
repeatedly confirmed.  In particular  \citet{SpiteMaSp84}, \citet{RyanNoBe99},  \citet{BonifacioPaSpetal2002}, \citet{Thorburn94}
 and \citet{Bonifacio2002} have confirmed 
a plateau at\footnote{$A(\Li) \equiv \log[N(\Li)/N(\H)]+12$.} $ A(\Li)\simeq 2.1$ or at $ A(\Li)\simeq 2.3$ depending on the \teff{} scale used.
$A(\Li)=2.32$ has been obtained by \citet{BonifacioPaSpetal2002} in turnoff stars of  NGC 6397, a low metallicity cluster with $[\Fe/\H]=-2.01$. 
\emph{While the Li concentration is nearly
the same in stars whose  Fe concentration differs
by more than a factor of 100}, from [Fe/H] $= -3.7$ to $-1.5$ \citep{Cayrel98}, a small progressive increase with [Fe/H] has perhaps 
been detected by \citet{Thorburn94} and \citet{RyanNoBe99}.

 The $ A(\Li)\simeq 2.1$ value  follows if one uses the lower \teff{} 
scale preferred for instance by \citet{RyanNoBe99} whereas the $ A(\Li)\simeq 2.3$ value follows if
one uses the high \teff{} scale for dwarf stars \citep{GrattonCaCa96}.  Partly because of the discussion in Sec. 3 of \citet{VandenBergRiMietal2002}
and the better agreement with low metallicity cluster isochrones shown in that paper if the high \teff{} scale is used,
we tend to prefer that scale.  
Consistancy between field and cluster stars also favors the high \teff{} scale.
We will however do comparisons of models to observations using both values of $A(\Li)$.

The results of 3D$-$NLTE simulations confirm the 1D$-$LTE Li abundance determinations in low metallicity 
Pop II stars \citep{AsplundCaBo2003} though changes to collision rates used in NLTE calculations might 
slightly lower the measured Li abundance value
\citep{BarklemBeAs2003}.

The Big Bang produces $^7\Li$ but no $^6\Li$.  $^6$Li is currently believed to be the product of
 galactic cosmic rays (GCR) \citep{Reeves94}.  However any observation of $^6\Li$ puts stringent constraints 
on internal transport processes in stars and observations of $^6\Li$ so shed light on $^7\Li$ observations.
$^6$Li  is more fragile than $^7$Li in that it is destroyed at a lower $T$ than $^7$Li by nuclear reactions.
If $^6$Li has survived, very little $^7$Li may have been destroyed by nuclear reactions.
Its surface abundance  is consequently another marker of internal turbulence.   
Its presence has been confirmed in a few Pop II field stars. In low metallicity stars,
\citet{NissenAsHietal2000} obtained the  ratio $^6$Li/$^7$Li$= 0.02 \pm 0.01$ in G 271-162.
\citet{SmithLaNi98} obtained  the  ratio $^6$Li/$^7$Li$= 0.05 \pm 0.03$ in BD $+26^0 3578$  and $^7$Li/H$=1.7 \times 10^{-10}$.
In HD 84937, \citet{CayrelSpSpetal99}  obtained  the  ratio $^6$Li/$^7$Li$= 0.052 \pm 0.019$ and $^7$Li/H$=1.6 \times 10^{-10}$;
 they also note that $^6$Li appears to be present only in the turnoff (or just past turnoff) halo stars.
This suggests that, at least in these three turnoff halo stars, very little surface $^7\Li$ has been destroyed.  Their surface $^7$Li
is then the one from which the star formed or it may have been reduced only by atomic diffusion.

Since $^6\Li$ is made by GCR, its presence in a star implies that some of the original $^7\Li$ in that
star also came from GCR and so
from another source than the Big Bang.  Since $^6\Li$ was measured only in stars with $[\Fe/\H] \simeq -2.4$, 
one estimates\footnote{Since in those stars where it was measured, $^6\Li$ is 
about 1/20 of the $^7\Li$ and since GCR produce approximately 1.5 times
more $^7\Li$ than $^6\Li$ according to \citet{Reeves94}.} that all stars with that metallicity had a $+0.06$ to $+0.1$ dex contribution to  $^7$Li/H
(see for instance \citealt{RyanNoBe99}). 
Since $^6\Li$ is the product of galactic production, its abundance increases at the same time as that of Fe, 
though the relation  between the two depends on the galactic nucleosynthesis model used.  That it should have been  observed
in turnoff stars only, appears to be consistent with stellar evolution models stating that it is more 
easily destroyed in cooler stars than in turnoff stars (see below).
One then expects a 0.06 to 0.1 dex contribution to $A(\Li)$ from GCR in all stars with $[\Fe/\H] \simeq -2.4$ and $\teff{} \simeq 6300$ K,
so that their $A(^7\Li)$ from cosmological origin becomes 2.0~dex in the low 
\teff{} scale and 2.2 dex  in the high \teff{} scale.
The high \teff{} scale gives a factor of 2.4 $\pm 0.4$ decrease from the original cosmological value (Eq. \ref{eq:wmapli}).  Given  
the various uncertainties, 
we will consider both a factor of 2.0  and of 3.0 as the reduction factor from the original cosmological $^7\Li$ to that observed today.

\section{Evolutionary model predictions}
\label{sec:evolution}

\subsection{Two series of models}
\label{sec:models}

For the first series of models, details of the calculations may be found in  \citet{TurcotteRiMietal98}, \citet{RichardMiRi2001}
 and \citet{RicherMiRoetal98}.
All effects of atomic diffusion are taken
into account.  Only quantities determined from first principles are used except for the mixing length parameter, $\alpha$,
which is calibrated using the Sun  \citep{TurcotteRiMietal98}.
These are then the first self consistent models calculated from first principles.
This  was made possible by the recent availability of large atomic data bases
that include the data needed to calculate radiative accelerations, \gr{},
throughout stellar
models \citep{IglesiasRo91,IglesiasRo93,IglesiasRo95,IglesiasRo96,RogersIg92,RogersIg92a,Seaton93}.
Using this data, the \gr{} and Rosseland opacities are continuously calculated during evolution as the
relative concentration of species changes \citep{RicherMiRoetal98}. They play a major role in the particle transport
equations.  The formalism of \citet{Burgers69} is used to calculate
the transport velocities leading to 56 (28 chemical species and 2 equations per species) non-linear coupled
differential equations.
In this series, evolutionary models 
 have been calculated for Pop II stars of 0.5 to 1.2 \Msol{} with [Fe/H]$_0$ from 
$-4.31$ to $-0.71$ \citep{RichardMiRietal2002,RichardMiRi2002,VandenBergRiMietal2002} and for solar metallicity clusters 
by \citet{MichaudRiRietal2004}. These models are called the \emph{models with atomic diffusion} in what follows.
In this paper, we mainly use models with $[\Fe/\H]=-2.31$.

The second series of models is similar to the first with $[\Fe/\H]=-2.31$, except that a turbulent transport coefficient, $\Dturb$, is added.
Models for this series are described  in \citet{RichardMiRietal2002}  but additional models were
 calculated for this paper.  Our aim in calculating models with turbulence is to
determine the level of turbulence that the $^7\Li$ and $^6\Li$ observations require.
This may later be used to determine the physical mechanism causing  this turbulence.  
These models are called the \emph{models with turbulence} in what follows;  in addition to turbulent diffusion \emph{they also 
include all effects of atomic diffusion}.

\subsection{Depth of convection zones and turbulence parametrization}
\label{sec:ZC}

\begin{figure}[H]
\centerline{
\includegraphics[width=0.8\textwidth]{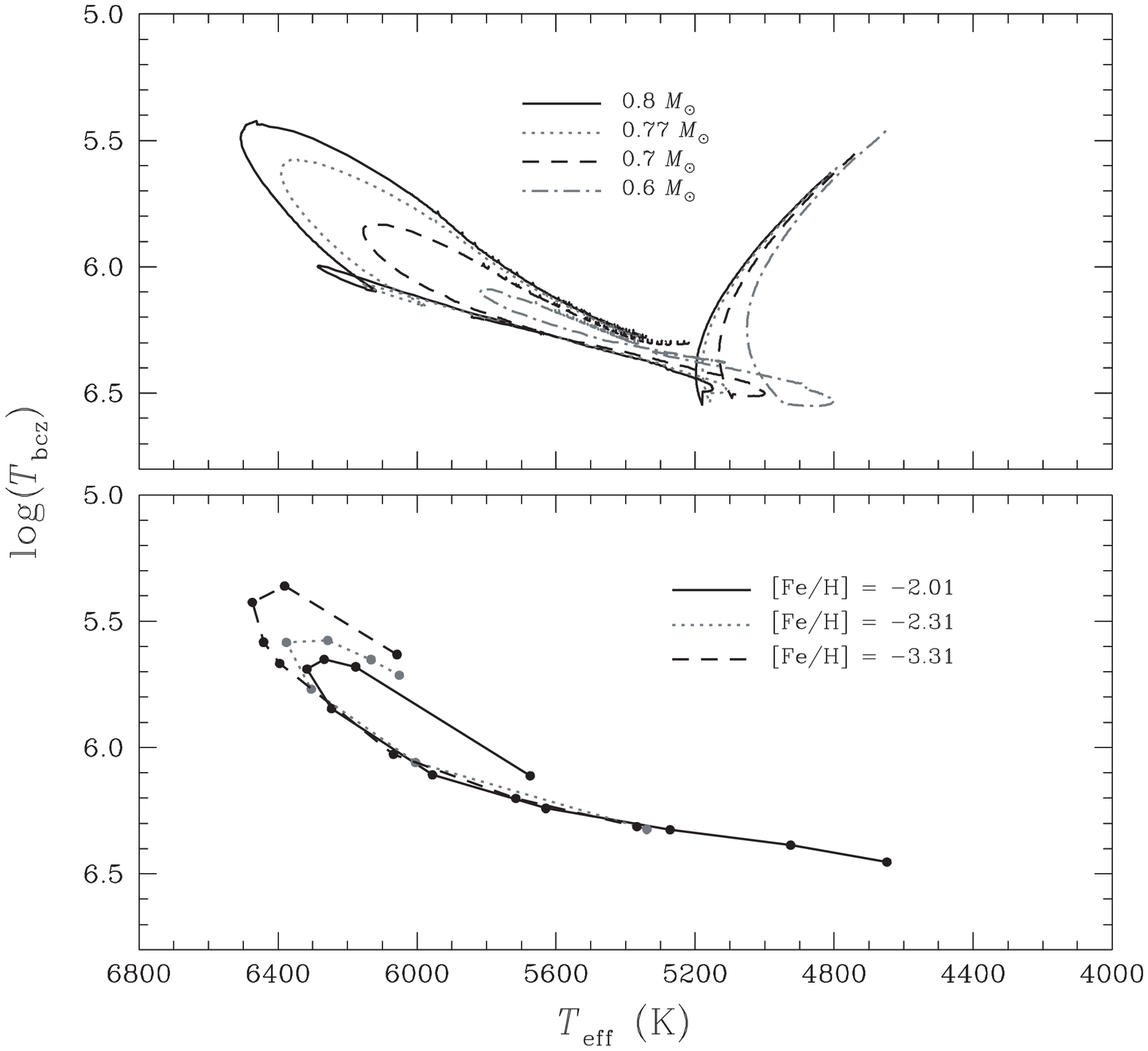}}
\caption{Temperature at the bottom of the surface convection zone as a function of  \teff{}.
In the upper part, are shown the paths followed during the evolution of stars of 0.6, 0.7, 0.77 and 0.8 \msol{}
with $[\Fe/\H]=-2.31$.  Evolution starts on the upper right hand section  of each curve.  In the lower part 
of the figure are shown \tbcz{} isochrones for three metallicities at 13.5 Gyr.
The 0.77 \msol{} star is at turnoff on the $[\Fe/\H]=-2.31$ isochrone.  
At 13.5 Gyr, all \MS{} stars of a given \teff{} have the same $T$
at the bottom of their surface convection zone, \tbcz{}, irrespective of their metallicity.  This is not the 
case past turnoff however.  
}
\label{fig:dtbcz}
\end{figure}

For surface Li abundance variations, perhaps the most important property of  evolutionary 
models is how  temperature (see Figure \ref{fig:dtbcz})
and mass (see Figure~1, 2 and 12 of \citealt{RichardMiRietal2002}) vary at the base of the surface convection zone.  
The temperature, \tbcz, is directly related to  Li burning while the mass, \mbcz, is directly related 
to  the time scale of gravitational settling (see \citealt{Michaud77a}).
Even in one single model, say the 0.77 \Msol{} one, the \mbcz{} decreases  by a factor of 30 during  \MS{} evolution
while \tbcz{} decreases by a factor of 3.  Most of 
the gravitational settling should occur around turnoff when \mbcz{} is smallest.
Because of the high sensitivity of Li burning to $T$, any  burning 
of  surface Li should occur early in the star's evolution when \tbcz{} is largest.
As will be shown in \S \ref{sec:diffusion} the reduction of surface Li does not necessarily occur through burning however.

On the lower part of Figure \ref{fig:dtbcz},  \MS{} stars are on the lower branch and stars past turnoff are on the upper 
branch of each curve.  An important characteristic is the independence of the \tbcz{} on metallicity during \MS{}
evolution.  It depends only on \teff.  This makes it possible for  processes that reduce the Li abundance to be independent
of [Fe/H].
 
Results will be shown for two different parametrizations of turbulent diffusion coefficients.
 One parametrization  allows to determine the turbulence which  minimizes the reduction of  surface $^7\Li$ abundance throughout the evolution
of stars of different masses. 
 The turbulent diffusion coefficient is defined 
  as a function of $T$ in order to  allow adjusting it 
to limit gravitational settling of $^7\Li$ while not burning $^7\Li$. 
 It is then essential
 to link turbulence to  $T$ since the rate of the nuclear reaction $^7\Li(p,\alpha)^4\He$ is
 highly $T$ sensitive; the Li burning occurs  at $\log T \simeq 6.4$ (see \citealt{LumerFoAr90} for a detailed discussion).  
To minimize Li abundance reduction one then tries to adjust the turbulent diffusion coefficient  to be smaller than atomic diffusion 
slightly below that $T$ so that turbulence reduces settling as much as is 
possible in surface layers without forcing $^7\Li$ to 
diffuse by turbulence to $\log T \simeq 6.4$.  
The parameters specifying turbulent transport coefficients are
indicated in the name assigned to the model.  For instance, in the T6.0D400-3 model,
 the turbulent diffusion coefficient, $D\T$,  is 400 times larger
than the He atomic diffusion coefficient at $\log T_0 = 6.0$ and varying as $\rho^{-3}$ or:
\begin{equation}
  \label{eq:DTT}
 \Dturb=400 D_{\mathrm{He}}(T_0)\left[\frac{\rho}{\rho(T_0)}\right]^{-3}. 
\end{equation}
 To simplify writing, T6.0 will also be used
instead of T6.0D400-3 since all models discussed in this paper have the D400-3 parametrization.  The $\rho^{-3}$ dependence is suggested (see 
\citealt{ProffittMi91a})
by observations of the Be solar abundance today showing it is hardly smaller than the original Be abundance (see for instance \citealt{BellBaBa2001}).  This imposes a rapid decrease of the turbulent transport coefficient as $\rho$ or $T$ increases.
Examples of those turbulent diffusion  coefficients and how they compare to atomic diffusion coefficients 
are shown in Figure\,6 of \citet{RichardMiRietal2002}.
Series of evolutionary models were calculated  with T6.0, T6.09, T6.25 and T6.28 in order to investigate the effect of various levels of  turbulence
and determine the parameters that lead to the observed level of surface Li abundance reduction.

The second parametrization of turbulent diffusion coefficients that is used here was introduced into 
  their solar model by \citet{ProffittMi91a}.
In this parametrization $\Dturb$ has always the same value immediately below the convection zone.  It is, in this model, implicitly assumed that  turbulence is generated from
motions in the convection zone.
\citet{ProffittMi91a} determined as a maximum level of turbulent transport 
allowed by the observed solar Li abundance: 
\begin{equation}
  \label{eq:DTS}
 \Dturb=7500\left[\frac{\rho}{\rho_{\rm{bcz}}}\right]^{-3}. 
\end{equation}
This model will be labeled PM7500.  When used in the solar model that serves as bench mark for  calculations of this paper, it reduces  $^7\Li$ 
at the solar age by a factor  $1.1 \times 10^{-3}$, whereas replacing 7500 by 4000 (not shown) and 2000 (labeled PM2000) in equation~(\ref{eq:DTS})
leads respectively to reduction factors of $1.2 \times 10^{-2}$ and $6.4 \times 10^{-2}$.  This was verified
by detailed stellar evolution calculations.
The observed solar photospheric \Li{} abundance is approximately $5 \times 10^{-3}$ 
what is thought to be the original abundance but a pre--\MS{} reduction by a factor of 3 to 15 is expected (see Table 1 of 
\citealt{PiauTu2002}) leaving a reduction of Li during the \MS{} by a factor of 15 to 70.
Solar models were calculated for this paper both with the PMx and Tx parametrizations.  In the case of the Sun, 
one can obtain the desired Li value at the age of the Sun with either parametrization.

\begin{figure}[H]
\centerline{
\includegraphics[width=0.65\textwidth]{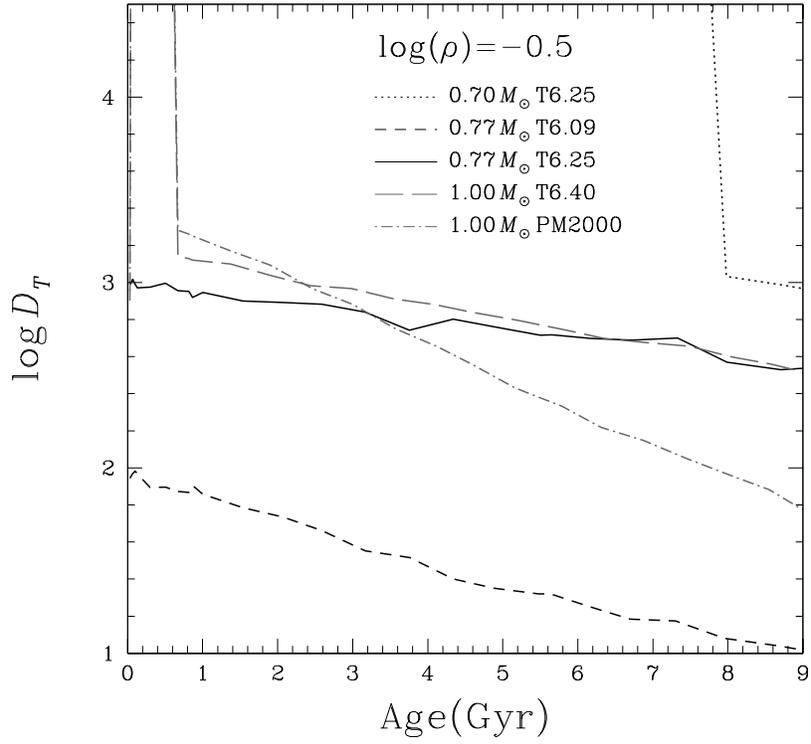}}
\caption{Turbulent diffusion  coefficient as a function of age at $\log \rho =-0.5$ in  0.7 and 0.77~\msol{} Pop II star 
and in the solar model.  The parameters are defined by Eq. (\ref{eq:DTT}) and (\ref{eq:DTS}) and the curves identified on the figure.  Bottom of convection zones are indicated by nearly vertical lines.
}
\label{fig:DT}
\end{figure}

The time evolution of the turbulent diffusion coefficients at $\log \rho =-0.5$  are compared
in Figure \ref{fig:DT}.  The density  $\log \rho =-0.5$ was chosen for the comparison because it 
is, during most of the evolution, between the convection zone and the region where Li burns
both in the solar and in the 0.77 \msol{} models.  Since for all calculations described in this paper
$\Dturb$ varies as $\rho{}^{-3}$, the comparison at one density is valid for all densities.

One notes from Figure \ref{fig:DT} that, at a given $\rho$, the $\Dturb$ of the 0.77$\msol{}$ T6.25 model is nearly equal that 
of the T6.4 solar model and of PM2000 throughout the evolution.  
In the  0.70$\msol$ T6.25 model it is a factor of 2.5 larger.  The 0.77$\msol$ T6.09 model has a factor of 10$-$30 
smaller $\Dturb$ than the other model.  
Using the PM2000 parametrization throughout the evolution of Pop II stars leads however to excessive Li destruction 
 in the lower mass objects.
The extent to which this is true will become more evident by considering Li isochrones in the next section.  
 The range of $\Dturb$ values
compatible with Li surface abundance in both the Sun and Pop II dwarfs is surprisingly small.

\subsection{Lithium isochrones}
\label{sec:lithium}
\begin{figure}[H]
\centerline{
\includegraphics[width=0.8\textwidth]{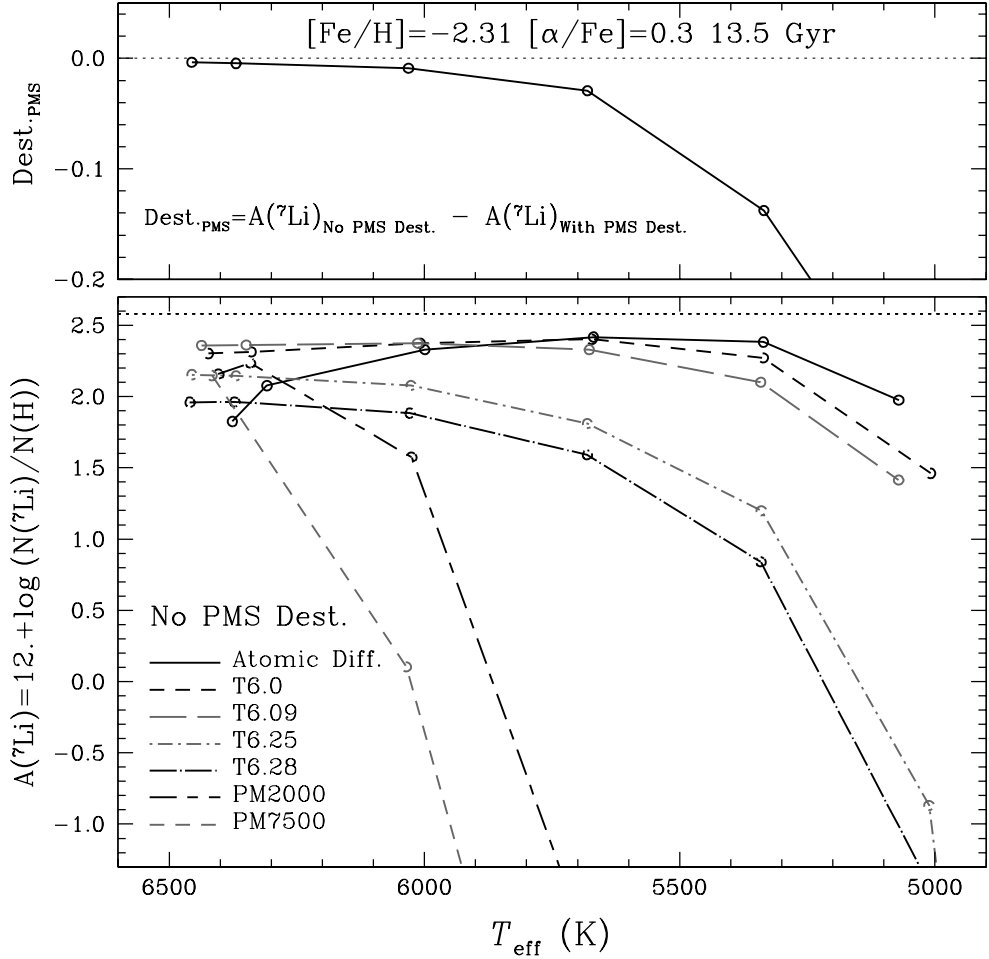}}
\caption{Isochrones of  $^7\Li$ at 13.5 Gyr in models where atomic diffusion only is included (continuous line) and where four different
levels of Tx turbulence are introduced.  Two isochrones for PMx models are also shown.
In the upper part of the figure is shown the pre--\MS{} destruction.  In the lower part is included the 
destruction during the \MS{} only. 
Open circles show the value calculated for the various models in stars of 0.77, 0.75, 0.7, 0.65, 0.6, 
and 0.55  \msol{} at 13.5 Gyr.
 No post turnoff star is included here  to avoid overcrowding.  See Figure \ref{fig:casino} where  post turnoff stars are included.
The horizontal dotted line identifies initial values.  
}
\label{fig:isoch7}
\end{figure}

\begin{figure}[H]
\centerline{
\includegraphics[width=0.8\textwidth]{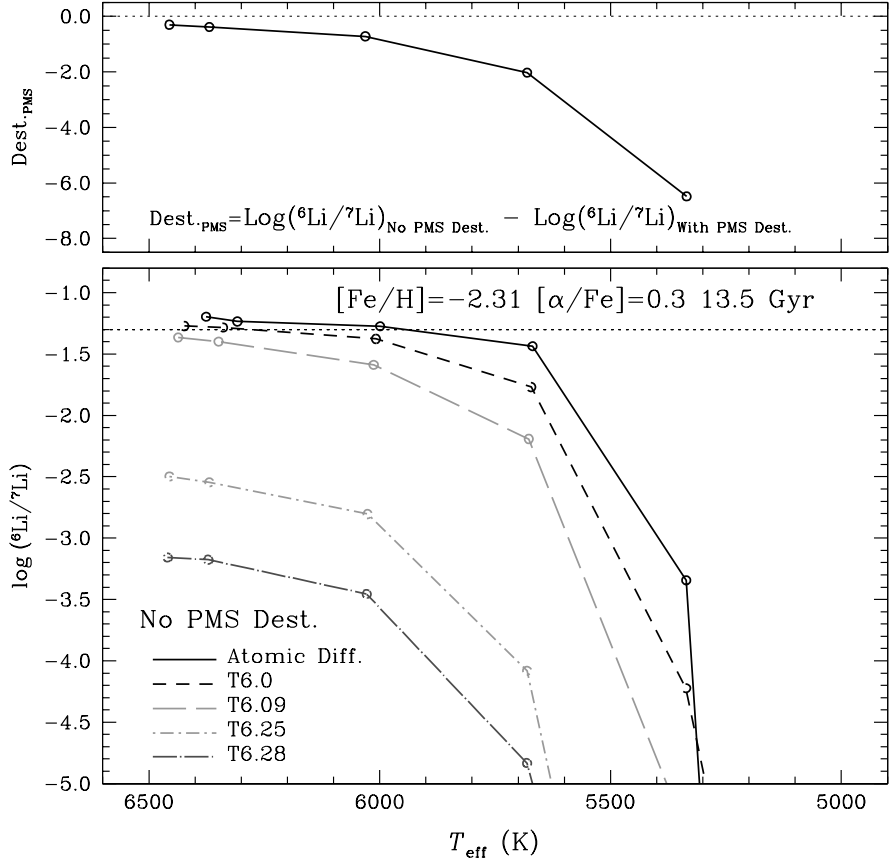}}
\caption{Isochrones of the $^6\Li$ to $^7\Li$ ratio at 13.5 Gyr in models where atomic diffusion only is included (continuous line)
 and where four different
levels of turbulence are introduced.  In the upper part of the figure is shown the pre--\MS{} destruction. 
 In the lower part, is included the 
destruction during the \MS{} only.   See the legend of Fig. \ref{fig:isoch7} for further details.
}
\label{fig:isoch6}
\end{figure}

Lithium abundance isochrones for  models with atomic diffusion and different series of models with turbulence are shown in Figures \ref{fig:isoch7} and \ref{fig:isoch6}.
Pre--\MS{} destruction is  independent of the level of turbulence since during the pre--\MS{} the whole star is mixed by a convection zone.
It is indicated separately in the upper part of the figures and is not included in the lower part.  The reduction factor on the pre--\MS{}
multiplies the reduction factor on the \MS.

On the \MS{}, the  $^7\Li$ abundance drops by at least $0.2$~dex in any star whatever the turbulence.
The situation is a little complex at  turnoff in that, while the $^7\Li$ abundance
is smaller (the reduction factor is larger) in the   model with atomic diffusion 
than in those with turbulence, at $\teff{}=6000$\,K 
the $^7\Li$ abundance
is smaller in most of the models with turbulence and decreases as turbulence is increased.  
At $\teff{}=6000$ K,  the model with T6.0 turbulence
has a very slightly larger  $^7\Li$ abundance than the  model with atomic diffusion.

 Also shown in Figure \ref{fig:isoch7} is the  isochrone calculated with turbulence linked to the position of
 the convection zone given by  equation (\ref{eq:DTS}),  labeled PM7500, and the PM2000 one. They both produce a large variation of Li.
 This parametrization  extends  the mixed zone below the bottom  of the convection zone
 by a fixed factor, large enough to minimize
 settling around 13.5 Gyr for instance in the 0.77 \Msol{} star (a factor of 100 or so is needed). This leads 
to complete destruction of
 Li in the early evolution of the stars with $M < 0.7\Msol$.

There has been  virtually no pre--\MS{} destruction of $^7\Li$ in PopII stars with $\teff{}> 5500$ K 
(Fig. \ref{fig:isoch7}).
 Pre--\MS{} destruction of $^7\Li$ is important for $\teff{} \lta 5400$ K.  
The level of destruction on the pre--\MS{} is  relatively uncertain.
\citet{ProffittMi89}   concluded that pre--\MS{} Li destruction was a sensitive function of a number of parameters 
such as metallicity and He content (see also 
\citealt{PiauTu2002}) and details of the convection model \citep{DantonaMo2003}.  \citet{PiauTu2002} obtain a factor of 20
reduction of surface $^7\Li$ during the solar pre--\MS{} (see their Fig. 10).  Pre--\MS{} destruction of $^6\Li$ 
is large for $\teff{}  \lta 6000$ K (see Fig. \ref{fig:isoch6}).  
This limiting \teff{} is uncertain but pre--\MS{} nuclear destruction appears unlikely to 
play a role at turnoff.

The original $^6\Li$ abundance (Fig. \ref{fig:isoch6}) is arbitrary and  chosen to reproduce the 
observed values in the turnoff stars
 where it was observed.  
The $^6\Li/^7\Li$ ratio is reduced by a factor of about 4 on the pre--\MS{} at 6000 K and by 
very large factors at smaller \teff{} so that one should expect to observe $^6\Li$ only above 6000 K. 
$^6\Li$ is less affected by atomic diffusion than $^7\Li$ so that \emph{the ratio $^6\Li/^7\Li$ is larger than  initially at  turnoff
in stars where only atomic diffusion is important for both $^6\Li$ and $^7\Li$}.  In both the T6.25 and T6.28 models $^6\Li$ is destroyed 
by large factors even in turnoff stars.

One important characteristic of the isochrones is that, at a given \teff{} before turnoff, the Li abundance  does not depend on metallicity.  
This is related to  
the property of  \mbcz, shown in Figure 2 of \citet{RichardMiRi2002}, and of \tbcz{} shown in Figure \ref{fig:dtbcz}  above.  Throughout \MS{}
evolution, \mbcz{} and \tbcz{} depend only on \teff{} independent of metallicity over the metallicity range of interest. 
 Consequently, for the small metallicities considered,  settling depends only on \teff{}
and the distance between the \mbcz{} and the temperature where Li burns depends only on \teff{}.  
This property may be seen, for the models with atomic diffusion, in Figures 9 and 11 of \citet{RichardMiRi2002}.  
After turnoff variations are seen as a function of metallicity however.
That the Li abundance reduction at a given \teff{}
is independent of metallicity before turnoff is then easy to understand.  What is more surprising is that there should be a 
relatively wide plateau as a 
function of \teff{}.

\subsubsection{Effect of He abundance variations}
\label{sec:he_iso}
At the same time as Big Bang models predict  a \Li{} abundance from WMAP observations, 
they also predict  an original \He{} mass fraction of $Y_{\rm{init}}= 0.2484$ which is slightly larger 
than $Y_{\rm{init}}= 0.2352$  used in all calculations presented in this paper except in this section. 
 The original He abundance used for most calculations 
is the same as used in \citet{RichardMiRietal2002}, \citet{RichardMiRi2002} 
and \citet{VandenBergRiMietal2002}.
This choice was made partly for consistency with previous calculations and partly because we had a large number 
of calculations available with that \He{} abundance.  
Furthermore previous results have shown that such a small \He{} abundance change 
had little effect on the turnoff luminosity
vs age relation (see Fig. 4 of \citealt{VandenBergBoSt96}).

\begin{figure}[H]
\centerline{
\includegraphics[width=0.8\textwidth]{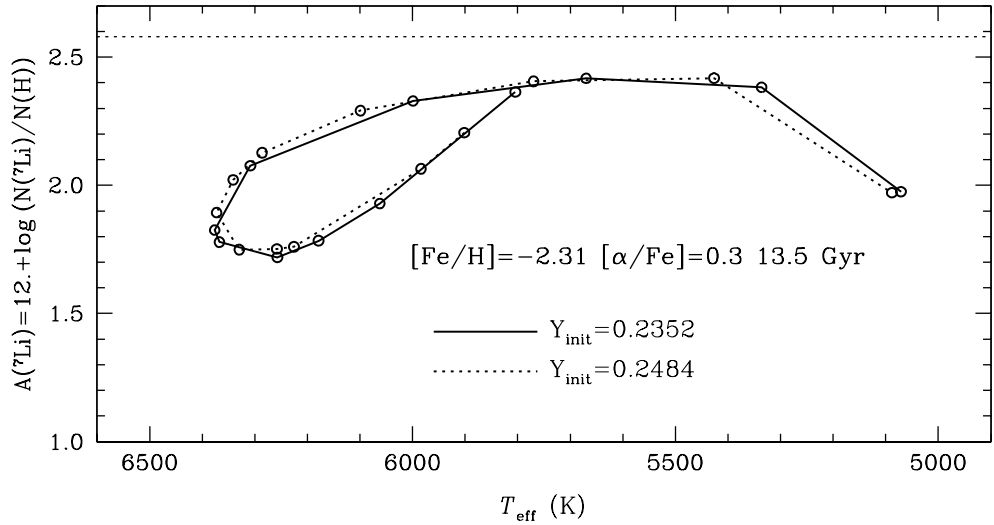}}
\caption{Isochrones of  $^7\Li$ at 13.5 Gyr in models where atomic diffusion only is included  and where the 
original \He{} abundance is varied.
The continuous line is the same as in Fig. \ref{fig:isoch7} and was calculated with the same original He mass fraction
of 0.2352 whereas the dotted line was calculated for a slightly larger original \He{} mass fraction.  The effect is negligible
and the difference between the two curves is mainly caused by the use of straight line segments
between calculated points. 
}
\label{fig:isochHe}
\end{figure}

One may however wonder if the quantities that depend most sensitively on the value of the \mbcz{}, 
such as the surface \Li{}  isochrones may not be affected by such a change in \He{} initial abundance.
We have calculated a series of models with diffusion with $Y_{\rm{init}}= 0.2484$ in order to determine the
size of the effect. \Li{} abundance isochrones calculated in both the diffusion models calculated
with $Y_{\rm{init}}= 0.2352$ and those with  $Y_{\rm{init}}= 0.2484$ are compared in Figure  \ref{fig:isochHe}.
The differences are very small and are mainly due to our use of straight line segments between calculated points. 
In stellar models of a given mass, the differences are significant but at a given age and \teff{}, the differences
in surface \Li{} abundance are negligible.  The main change is in the mass of the model that is at a given evolutionary
step.
At 13.5 Gyr  the $Y_{\rm{init}}=0.2484$ model with maximum \teff{} has  $\Mstar{}= 0.75 \Msol$, whereas 
 the model with $Y_{\rm{init}}=0.2352$ with maximum \teff{} has 0.77 $\Msol{}$.  The models with the minimum 
values of \Li{}
have respectively masses of 0.76 and 0.78 $\Msol$.  So at a given age, the isochrones are very nearly the same 
but for different mass models.

In Figure  \ref{fig:isochHe}, we have continued the isochrones on the subgiant branch where the 
surface \Li{} abundance is most sensitive
to the \mbcz{}.  This was not done on Figure \ref{fig:isoch7} in order to avoid confusion with other curves.

We have also verified (not shown) that the (\teff{}, $L$) isochrones were not significantly modified, confirming the
analysis of \citet{VandenBergBoSt96}.

\subsection{Settling vs destruction of surface Li}
\label{sec:diffusion}

To understand what the constancy of the Li abundance as a function of \teff{} implies, one needs to understand
 the transport mechanisms below surface convection zones. 
Nuclear burning, atomic diffusion and turbulence are involved.

\subsubsection{Nuclear burning}
\label{sec:nuclear}

From an analysis of nuclear reaction rates and a comparison to detailed calculations,
\citet{ProffittMi89} obtained that (see their Eq.\,1) the relationship between the remaining  fraction of
the two Li isotopes  may be expressed by:
\begin{equation}
  \label{eq:6_7}
\frac{^6\Li}{^6\Li_0}=\left(\frac{^7\Li}{^7\Li_0}\right)^r.
\end{equation}
 The value of $r$ comes from the ratio of the nuclear reaction rates
and varies from  $r=84$ to 94 depending on the value of $T$ where most $^6\Li$ burns.
\citet{ProffittMi89} obtained equation (\ref{eq:6_7}) in pre--\MS{} stars where there is 
nearly instantaneous mixing of the whole star by  convection. During \MS{} evolution, 
the situation is more complex.

By comparing the isochrones for $^7\Li$ (Fig. \ref{fig:isoch7}) with those for the  $^6\Li$ to $^7\Li$ ratio (Fig. \ref{fig:isoch6}), it is
clear that the  abundance of $^6\Li$ varies much more slowly compared to that of $^7\Li$ than is given by equation (\ref{eq:6_7}) with $r=84$.
For instance at turnoff with T6.25 turbulence the   $^6\Li$ to $^7\Li$ ratio is reduced by about 1.2 dex while $^7\Li$ is reduced by
about 0.4 dex whereas equation (\ref{eq:6_7}) would have led to more than 30  dex reduction of $^6\Li$ for that reduction of $^7\Li$.  
This leads us to investigate the various contributions   to  atomic and turbulent diffusion velocities 
immediately below surface convection zones.
Burning occurs a finite distance from the bottom of the surface convection zone and the 
surface Li is destroyed only in so far as it is transported to the burning region 
by turbulence (see \citealt{VauclairVaScetal78} for  approximate evaluation formulas).  
The reduction rate of the surface $^6\Li$ and $^7\Li$ abundance depends   
on the transport efficiency from the bottom of the surface convection zone to the layer 
where $^6\Li$ and $^7\Li$ burn.  It also depends on how much of the surface abundance reduction
is due to atomic diffusion and how much to turbulent transport to the burning region.

\subsubsection{Settling vs turbulent transport}
\label{sec:settling}

While, in our evolutionary models, a complex set of 56 coupled differential equations is used,
 in order to interpret the 
results of the evolutionary models, it is convenient to use 
a  diffusion equation developed for ternary mixtures by 
\citet{AllerCh60}. One must add to that equation the differential radiative accelerations, \gr{},
 as introduced by \citet{Michaud70} as well as a turbulent diffusion term, $\Dturb$,
as introduced by \citet{Schatzman69}\footnote{Turbulence is made up of a wide spectrum of macroscopic advective motions, 
but Schatzman showed they could be modeled by a diffusive term.  Anisotropic turbulence is slightly 
more complex to model (see \citealt{VincentMiMe96}).}:
\begin{equation}
v_D=(D_{ip}+\Dturb)\left[  - \frac{\partial \ln c_{i}}{\partial r}\right] + D_{ip} \left[ \frac{A_i m_p}{kT}(\gr{}_{,i}-g)+\frac{Z_i m_pg}{2kT} +k_T\frac{\partial \ln T}{\partial r}\right].
\label{eq:vDT}
\end{equation}
Within the  first brackets on the right is the purely diffusive term which includes a contribution both from atomic diffusion, $D_{ip}$,
 and from turbulent diffusion, $\Dturb$.  Within the second brackets on the right,  is an advective term caused by radiative acceleration, 
gravity,  electric field, and thermal diffusion.  In the \emph{ models with atomic diffusion}, $\Dturb= 0$.  In the models with turbulence, 
the parametrization used for \Dturb{} is defined 
by equations (\ref{eq:DTT}) and (\ref{eq:DTS}).  The $\gr{}_{,i}$ dominate all other terms in turnoff low metallicity stars for Be, B and metals 
(see \citealt{RichardMiRi2002,RichardMiRietal2002}).

In Figure \ref{fig:vdif} are shown separately the diffusive (dashed) and the advective (dot-dashed) 
components\footnote{The diffusive component shown here was calculated in post processing using Eq.\,(\ref{eq:vDT})
 and might differ slightly from the more accurate value
 used by the evolutionary code. The advective component was available in the code output and the one shown 
is the same as used by the code.}
 of the  $^6\Li$ and $^7\Li$ diffusion velocities as well as the total velocity (continuous) at an age of 13.5 Gyr in a 0.77\msol{} model 
with T6.25 turbulence. Immediately below the convection zone, the diffusive part of the velocity, $v_c(^7\Li)$,
is \emph{towards the surface} for $^7\Li$ while $v_c(^6\Li)$ is towards the interior but
 equals only $1/4$ of the total downwards velocity  for $^6\Li$. 
The advective part of the 
diffusion velocity is towards the interior for both isotopes and causes the reduction of the $^7\Li$ abundance and most 
of that of $^6\Li$ in the surface convection zone.
The diffusive part of the velocity is mainly driven by turbulence.  When it is toward the interior, it is
linked to internal nuclear burning.  Immediately below the  convection zone, it is however towards the surface for $^7\Li$
and is reacting to Li settling being larger close to the surface.  The diffusive term is then reducing 
the effect of gravitational settling, 
just as it does reduce the effect of the advective term for other metals.  However for Li, it has only a  
limited effect because nuclear reactions have virtually eliminated Li deeper in.  The corresponding Li profiles
 are shown in Figure \ref{fig:prof_li}.   There is only a very small 
buffer which turbulence can use to maintain the surface abundance.

 The $^7\Li$ profile for the T6.25 model is very nearly horizontal
for $\log T < 6.2$ (Fig. \ref{fig:prof_li}).  The transport due to the diffusive term is a large fraction of the total transport velocity 
and changes sign  (at 
$\log T \sim 5.9$) but the 
Li abundance appears constant because the \Dturb{} used are very large so that a very small Li abundance gradient is sufficient
to cause the purely diffusive term of the velocity.  The burning $^7\Li$ at $\log T \sim 6.4$ 
forces the downward diffusive term for $\log T \gta 5.9$
but is unable to affect more superficial regions.  The diffusive term is upwards for $\log T< 5.9$.

In the absence of turbulence (continuous line in Fig. \ref{fig:prof_li}), the sign change of the slope of the Li profile is 
well defined at $\log T\sim 6.3$.
The purely diffusive term leads to a downwards velocity for  $\log T> 6.3$  but to an upwards 
velocity for $\log T< 6.3$ (not shown).  The purely diffusive term of the T6.0 and T6.09 models changes sign also at $\log T\sim 6.3$
but, because $\Dturb + D_{ip}$ is very large compared to $D_{ip}$ alone, 
the Li abundance gradient is very small for $\log T < 6.15$.  Note that the total diffusion velocity (including both
diffusive and advective components) is always downwards.

The $^7\Li$ burning causes a sink at $\log T \sim 6.4$.  In the  model with atomic diffusion, the diffusive term causes the
burning to extend its effect to $\log T \sim 6.3$.  The advective terms in equation (\ref{eq:vDT}) cause a reduction of the 
surface Li by a factor of $-0.7$ dex (see Fig. \ref{fig:prof_li});  the surface underabundance is limited by the upwards
diffusion caused the purely diffusive term.  If T6.0 turbulence is present, the Li burning is hardly affected but
the upwards diffusive term is increased and cancels the downwards advective term for a very small abundance gradient
limiting surface underabundance to $-0.22$ dex.  With T6.09 turbulence, it is limited to $-0.19$ dex again with
little additional burning.  As turbulence is further increased
to T6.25, additional burning occurs however.

Atomic diffusion is then largely responsible for  the transport
 of $^6\Li$ and $^7\Li$   immediately below the surface convection zone not only in the 
model with atomic diffusion but even in the T6.25 model.
 The transport  by atomic diffusion 
 determines the reduction  of $^6\Li/{^7\Li}$.  Even when turbulence reduces the effect of atomic diffusion 
for metals below 0.1 dex, atomic diffusion is still the dominant transport process for $^7\Li$ below the surface convection zone.  
We have also verified (not shown) that, in the absence of atomic diffusion, the surface $^6\Li/{^7\Li}$ abundance reduction
is dominated by turbulent transport and the reduction factor does not approach the value of equation (\ref{eq:6_7}) with
$r\approx 90$.  It is only when  one assumes instantaneous mixing between the Li burning region and the surface that 
the reduction factor approaches this value.  This is appropriate for instance on the pre--\MS{} \citep{ProffittMi89}.
Otherwise, the transport mechanism, be it atomic or turbulent diffusion, plays the dominant role in determining the 
surface reduction factor.

\section{Discussion}
\label{sec:discussion}

\subsection{Li abundance and upper limits to turbulent transport}
\label{sec:turbulent}

In Figure \ref{fig:casino}  calculated surface abundances are
 compared to  observations of Li in halo stars and in two globular clusters.
On the upper part of Figure \ref{fig:casino} is shown the calculated surface Li concentration (black open circles)
in 50 stars of initial metallicity  $[\Fe/\H]=-2.31$.  These are the result of a Monte Carlo simulation based
on interpolations among a dozen complete evolutionary tracks (see Fig. 15 of \citealt{RichardMiRi2002} for the results of a different draw in which [Fe/H] was also allowed to vary).  No turbulence is assumed outside of convection zones.  
The age of stars 
was randomly generated around 13.5 Gyr with a gaussian distribution of 0.3 Gyr standard deviation; 
90 \% of the generated stars are between 13 and 14 Gyr.  Only stars with $\log g \ge 3.8$ are included.
Observations of Li abundance in metal poor halo stars
by \citet{Thorburn94}, in very metal poor stars by \citet{BonifacioMoSietal2003} and in the globular clusters M\,92 and NGC\,6397 
\citep{Bonifacio2002,BonifacioPaSpetal2002} are also shown.
The Li observations below 5500 K may have 
been affected by pre-main sequence evolution (see the upper part of Fig.~\ref{fig:isoch7})
which might explain the lower Li abundances observed in those stars.
However it does not appear possible to avoid, with our models with atomic diffusion, a progressive reduction 
of the predicted Li 
abundance as \teff{} increases above 6000 K.  
Two different values of Li abundance at a given \teff{} are expected above 6000 K since 
post$-$turnoff stars have smaller Li abundance than pre$-$turnoff stars.
A few stars with smaller Li abundance than plateau stars have been observed, and 
may be explained by these results (for instance the star represented by a magenta cross at $\teff{}= 6350$ K).
However no general reduction of Li abundance is  observed above 6000 K.
As in Pop I stars, additional turbulence appears to be required \emph{at least in some stars}
 \citep{RicherMiTu2000}.

Series of models with different assumptions about the strength of turbulence were 
then calculated.  One series uses T6.25 turbulence and the result
of a simulation for 50 stars is shown on the lower part of Figure~\ref{fig:casino}.
Observations of Li abundance in metal poor halo stars
by \citet{SpiteMaSp84}  and by \citet{RyanNoBe99} 
are also shown.  For \teff{} $ \ge 5600 $ K, the Spite plateau is very well reproduced.
This level of turbulent transport leads to a slight overdestruction of $^7\Li$ for $\teff{} \le 5600$~K.  

The hottest stars simulated with models with turbulence are some 150~K hotter than the hottest ones simulated
with models with atomic diffusion (see also \citealt{VandenBergRiMietal2002}).  If one uses the hotter \teff{} scale, 
as used for some of the observations reproduced on the upper part of the figure, the hottest stars observed have a \teff{}
compatible with that of the hottest stars simulated with models with turbulence; however if one  uses the
 lower \teff{} scale, 
as used for some of the observations
included in the lower part of the figure,  the hottest stars observed have a \teff{}
compatible with that of the hottest stars simulated  with  models with  atomic diffusion.
  This assumes an age of 13.5 $\pm 0.3$Gyr which is compatible with the one determined by WMAP.

Lithium isochrones of Figures~\ref{fig:isoch7} and  \ref{fig:isoch6} and the comparison to observations of 
Figure~\ref{fig:casino}  show that, if halo low metallicity stars 
 have  13.5  $\pm$ 0.3 Gyr:

1) The model with atomic diffusion leads to 0.8 dex underabundance at turnoff  or 0.5 dex more than
at 6000 K.  This  suggests the presence of some weak turbulence below the convection zone.

2)  With the T6.09 model, the  underabundance is approximately $-0.2$ dex in all stars within the \teff{} interval of 
the Spite plateau as discussed in \citet{RichardMiRietal2002}.  Figure~15 of that paper also shows that 
the high constancy of the Li abundance observations above 5600 K can be reproduced and even exceeded 
in a self consistent model with turbulence.

3)  The level of required turbulence depends on the \teff{} scale used to analyze the data.
Given the various error bars, T6.09 turbulence is probably not excluded.  However
if the low \teff{} scale used by  \citet{RyanNoBe99} is the right one, T6.25 turbulence would be required 
in turnoff stars.  The absence of observed stars above 6350 K (see Fig. \ref{fig:casino}) then poses a problem.
Turbulence should  be limited to lower $T$ in cooler stars than in turnoff stars 
since otherwise, there is excessive destruction below 5600 K.

4)  Turbulence appears linked to $T$ or to density more than to the bottom of the convection zone.
All  Tx models lead to fairly horizontal isochrones but with progressively larger 
Li abundance reduction as x increases. 
Depending on how close to the Li burning temperature the  turbulence extends determines the destruction factor.
The reason for the Li plateau in a given Tx model is relatively easy to understand.  In a given Tx model, the turbulence
extends to the same $T$ for all stellar masses  and so leads to similar burning flux in all stellar masses.  Since the 
mass above a given $T$ does not vary rapidly as a function of \Mstar{}, the Li reservoir to be burned is always about 
the same for all \Mstar{}.  At a given age, the remaining Li is then about the same independent of \Mstar{}.

5)  Turbulence that extends the convection zone by a fixed factor, such as given by equation (\ref{eq:DTS})
and which may be produced by overshooting, 
leads to completely  wrong \teff{} dependence of the Li isochrones. It overdestroys Li in  low \teff{}
stars.   It is not possible to extend convection zones by approximately a fixed level of turbulence 
\citep{ProffittMi91}.  The value given by equation (\ref{eq:DTS}) had been found by \citet{ProffittMi91a} 
to slightly overdestroy Li 
in their solar model.    It 
is an upper limit because part of the Li destruction is believed to have been caused by pre--\MS{} burning.
In the solar models of this paper, it however overdestroys Li by at least a factor of 10,
probably because of a deeper surface  convection zone as compared to the model of \citet{ProffittMi91a}.
One has to reduce the turbulence given by equation (\ref{eq:DTS}) by a factor of 3 to 4.  The model
compatible with surface solar Li abundance is that labeled PM2000 in Figure \ref{fig:isoch7}.

6)   It is of course possible to
determine a value of x such that Tx models have the observed Li destruction in the Sun since one has one
observation and one adjustable parameter. 
  The level of turbulent transport required to explain the solar value is around T6.40 or about 0.2$-$0.3 dex deeper in $T$ 
than required to reproduce the Li abundance of the Pop II Li plateau.  

 It is  however surprising  that the Tx required for the Sun should be so close to the 
Tx model that best reproduces Li in PopII stars.  As may be seen in Figure \ref{fig:DT}, the turbulent diffusion 
coefficient below convection zones in a 0.77 \msol{} T6.25 Pop II star is very nearly the same as in the T6.40 Sun throughout  evolution.

7)    Clearly there remains considerable uncertainty in the value of the turbulent diffusion
coefficients.   There are many potential sources of such turbulence  since, for instance, convective motions inside convection 
zones must lead to some turbulent motions beyond the frontier.  Differential rotation is another potential source.  
In fact what is surprising is how little turbulent extension of convection zones is allowed by Li abundance observations. 

8)  Surface $^6\Li$ is not destroyed by nuclear reactions on the \MS{} in models with 
atomic diffusion for $\teff{}  \gta 5500$ K.
In models with T6.09 turbulence, destruction of surface $^6\Li$ occurs only in stars with $\teff{}\lta 6000$ K.
However, in models with T6.25 and T6.28 turbulence, surface $^6\Li$ is destroyed by more than a factor of 
30 even in turnoff star.  \emph{The presence of $^6\Li$ in some turnoff stars appears to rule out T6.25 and T6.28 turbulence.}

9)  In all stars appearing in Figure \ref{fig:isoch6}, the ratio of $^6\Li$ to $^7\Li$ destruction is much
smaller than $10^{80}$.  A large fraction of the Li transport  has been caused by atomic diffusion in all
turbulence models considered (see \S \ref{sec:settling}).  We have also verified that even if one neglects atomic diffusion,
one still does not get the very large ratio given by equation (\ref{eq:6_7}).  Such a ratio applies only if $^6\Li$ 
destruction occurs \emph{in}
the convection zone.  As soon as a transport process is involved, the depletion factor depends more 
on the properties of the transport process than on ratios of nuclear reaction rates.  In most stars,  Li destruction occurs in the 
convection zone only while it is on the pre--\MS.

\subsection{Link to other work}
\label{sec:other}

The Li abundance in Pop II stars has generated a large number of papers and we do not
intend to review all of them.  It is however appropriate to analyze briefly the link to other points of view.

\citet{SalarisWe2001} argued that the current Spite plateau observations were consistent with 
evolutionary models including only atomic diffusion
if all plateau stars were 13.5 to 14 Gyr old and they also argued that this implied a reduction from the original 
abundance by a factor of about two, consistent with analyses of the cosmic microwave background.
To be consistent with observations as presented in Figure \ref{fig:casino}, this requires, as discussed by those authors,
uncertainties in the \teff{} scale conspiring to minimize the \teff{} dependence of the Li abundance on halo stars.
 The small sampling scale may also play a role and they argue
that more observations are needed to exclude, at an acceptable certainty level, that surface abundances are determined by stellar
evolution models with atomic diffusion without significant contribution from macroscopic mixing processes.
We agree that more observations of Li in halo stars are needed to firmly establish the origin of the Li abundance reduction 
and also that the \teff{} scale may well be off by some 100  or, perhaps, 200 K.  This may reduce the difference between the 
WMAP value and the Spite plateau to a factor of $\sim$ 2.  It seems more difficult to explain the continuity 
of the Spite plateau to the highest observed \teff{} without the presence of some weak turbulence in those stars.

The more frequently discussed point of view has followed  \citet{Vauclair88} analysis of the possible 
role of turbulent transport in  causing Li destruction and/or reducing Li settling.  Prescriptions
 for the initial angular momentum distribution, angular momentum loss, turbulent transport of angular momentum 
and of Li are required. Given the current level of understanding of turbulent transport processes, different
arbitrary parameters are used for angular momentum and particle transport even though efforts are made to link the two.  
Various prescriptions were studied but, given  input uncertainties, obtaining credible
models is a formidable task.
\citet{PinsonneaultDeDe91,PinsonneaultDeDe92} and \citet{ChaboyerDe94} obtained destruction factors of up to 10 and linked 
their assumptions of the effect
of internal differential rotation to observations of rotation on the horizontal branch.  
One has to assume an initial distribution of angular momentum and  one then tends to obtain excessive Li abundance dispersion
 \citep{VauclairCh95}.  
\citet{PinsonneaultWaStetal99,PinsonneaultStWaetal2002} strove to satisfy the small dispersion observed 
on the Spite plateau by varying input parameters and turbulent transport coefficients for Pop I and II stars. 
They were able to define a set of original angular momentum distribution 
and transport coefficients calibrations consistent with observed Li dispersion.  
While they do not explicitly say so, they appear to have neglected atomic diffusion in their calculations.  
\citet{PinsonneaultWaStetal99} concluded that the Li abundance depletion had to be between 0.2 and 0.4 dex.
Since little is known about either initial angular momentum, turbulent transport of angular momentum 
or turbulent transport of Li, one should perhaps not be too surprised that these studies have not been very conclusive.

 One difficulty with these models is that they neglect any potential role of internal magnetic fields in 
transporting angular momentum.  
In particular  models based on diffusive transport of angular momentum never reproduced the
 near solid body rotation of the Sun revealed by heliosismology \citep{CharbonneauToScetal98,CouvidatGaTuetal2003}. 
The flat rotation profile of the Sun is much more easily reproduced once 
magnetic fields are introduced \citep{BarnesMaCh99}.  As an alternative, \citet{TalonKuZa2002} have suggested that the
solar interior could be slowed down by internal gravity waves that might lead also to 
a Li Spite plateau \citep{TalonCh2004}.  Both the magnetic and gravity wave models have the advantage of being compatible with 
solar Be abundance observations while models which strive to explain momentum transport by turbulent diffusion 
 are not \citep{Balachandran2000}.

The difficulties with reproducing in detail the observations once transport processes are taken into account 
 led some to consider  \emph{standard} stellar
 evolution where it is \emph{arbitrarily assumed} 
that there is no 
transport process outside of convection zones so that no depletion from the original Li occurs on the Spite plateau
\citep{Chaboyer94a,CrifoSpSp98}.  There is no physical justification to assume that physical processes described from first
principles do not apply to the Spite plateau stars.

\section{Conclusion}
\label{sec:conclusion}

We have confirmed, using the latest evolutionary model calculations, the result of 
\citet{MichaudFoBe84} that atomic diffusion imposes a reduction by a factor of at least 1.6, and more probably 2.0,
 of the Li abundance observed in the oldest stars as compared to that produced in the Big Bang.   
As briefly described in \S \ref{sec:other}, this result 
is consistent with most other calculations done since, except 
\emph{when atomic diffusion
is arbitrarily neglected} in the calculations.  This result is also  confirmed observationally by the Li abundance  of the Spite
plateau being a factor of 2 to 3 smaller than the primordial value inferred from WMAP as described 
in \S \ref{sec:context} and \S \ref{sec:observations}.
Independent of the degree of turbulence in the 
outer regions, \citet{VandenBergRiMietal2002} (p. 496) have obtained for  M92 an age of 13.5 Gyr 
which is  consistent with the WMAP determination 
of the age of the universe and which is 2 Gyr smaller than the age determined 
by  \citet{GrundahlVaBeetal2000} who used models in which atomic diffusion was neglected.

The limitation of the approach comes from the constancy of the Li abundance in low metallicity Pop II stars 
above $\teff{}\simeq 5600$ K.  When special care is taken to reduce as much as possible the observational errors, 
as done for instance by \citet{RyanNoBe99}, one obtains a smaller dispersion for most stars with $\teff{} > 6000 $K.
  This is the reason for part of the difference between the observations on the upper and lower parts of 
Figure~\ref{fig:casino}.
Part of the scattering of the observations of \citet{Thorburn94} appears to be observational.  It appears 
that the star to star variation of the Li abundance is at most 0.1 dex at a given \teff{} except for a number of stars that 
are a factor of 2 or more below the plateau.  This excludes any process that would lead to large abundance variations and would depend 
on a property varying from star to star such as original stellar angular momentum, for instance.  
It is however consistent with atomic diffusion
being the main Li transport process from the convection zone, as found in \S \ref{sec:diffusion} to be the case even in presence of 
turbulent transport. Turbulent transport is then a perturbation to the main transport process from the convection zone which
is gravitational settling.

\begin{figure}[H]
\centerline{
\includegraphics[width=0.8\textwidth]{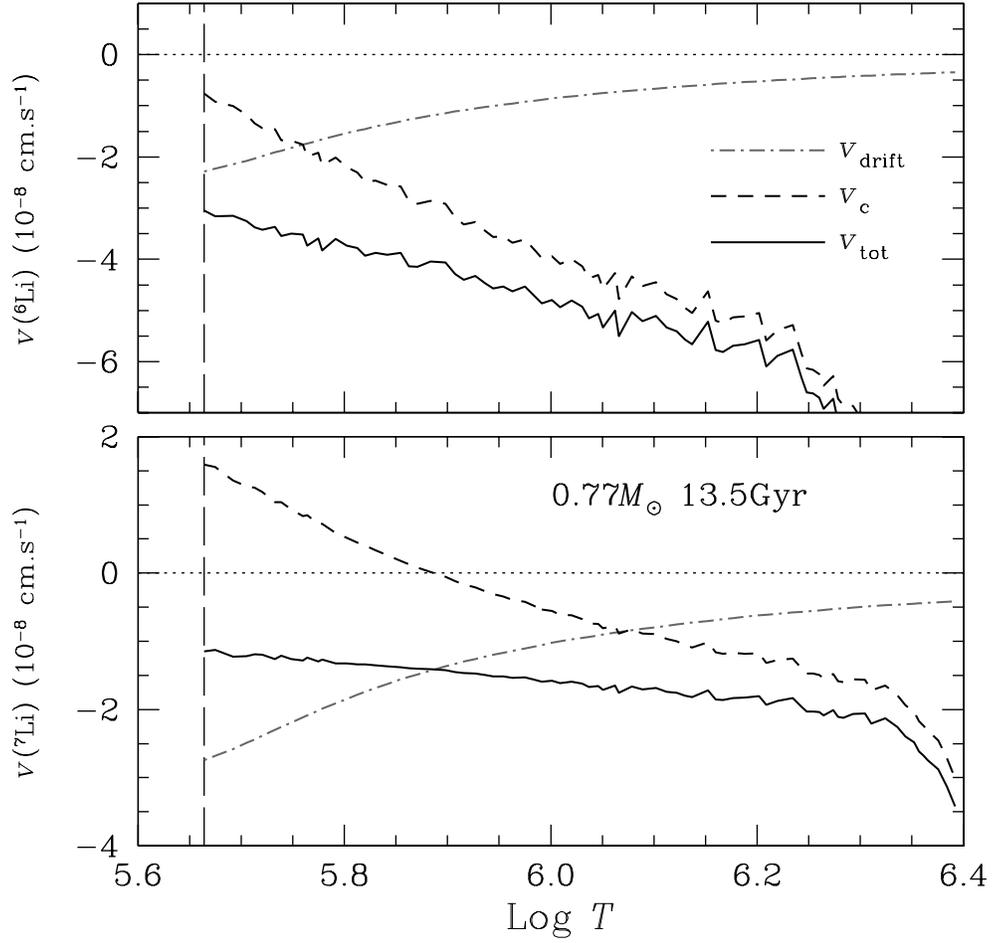}}
\caption{Diffusion velocities of $^6\Li$ and $^7\Li$ at 13.5 Gyr in a 0.77 \msol{} model with T6.25 turbulence. 
 The continuous lines are the total velocities, ${v_{\rm tot}}$, while the
dot-dash line is the advective velocity, ${v_{\rm drift}}$, and the dash line the purely diffusive term, ${v_{\rm c}}$, containing both atomic 
diffusion and turbulent diffusion.  Negative velocities are towards the interior. 
The bottom of the convection zone is indicated by a vertical long dashed line.   Immediately below the convection zone, 
the main inward contribution to the total velocity comes from 
the advective term and is caused by gravitational settling.  The dashed line was calculated in postprocessing
where only a limited number of significant figures was available leading to  the noisy character of those curves.
}
\label{fig:vdif}
\end{figure}

\begin{figure}[H]
\centerline{
\includegraphics[width=0.6\textwidth]{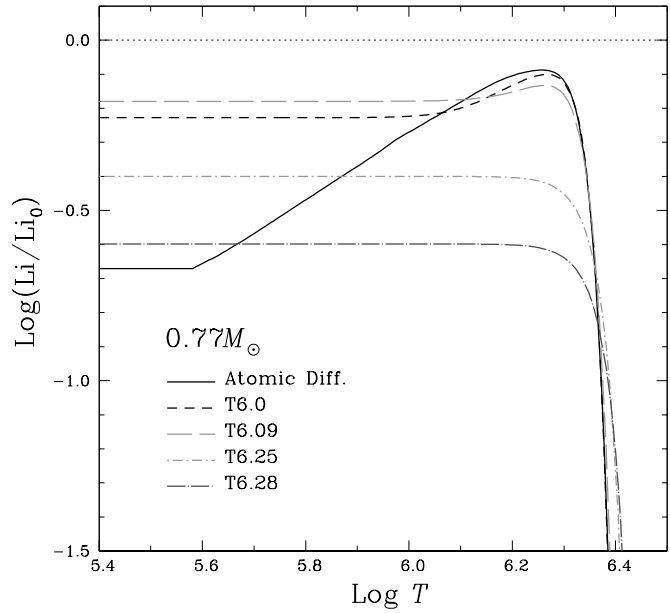}}
\caption{ $^7\Li$ profile in  0.77 \msol{} stars at 13.5 Gyr of $[\Fe/\H]=-2.31$ with various turbulence models.
The smallest surface $^7\Li$ abundance is in the model with atomic diffusion (continuous line).  As turbulence is increased, the surface
abundance first increases in the T6.0 and T6.09 models but then decreases as turbulence  is further increased in the T6.25 and T6.28 models.
On the other hand the maximum in Li concentration seen at $\log T \sim 6.25$ in the model with atomic diffusion
is progressively reduced as turbulence is increased.
}
\label{fig:prof_li}
\end{figure}

\begin{figure}[H]
\centerline{
\includegraphics[width=0.85\textwidth]{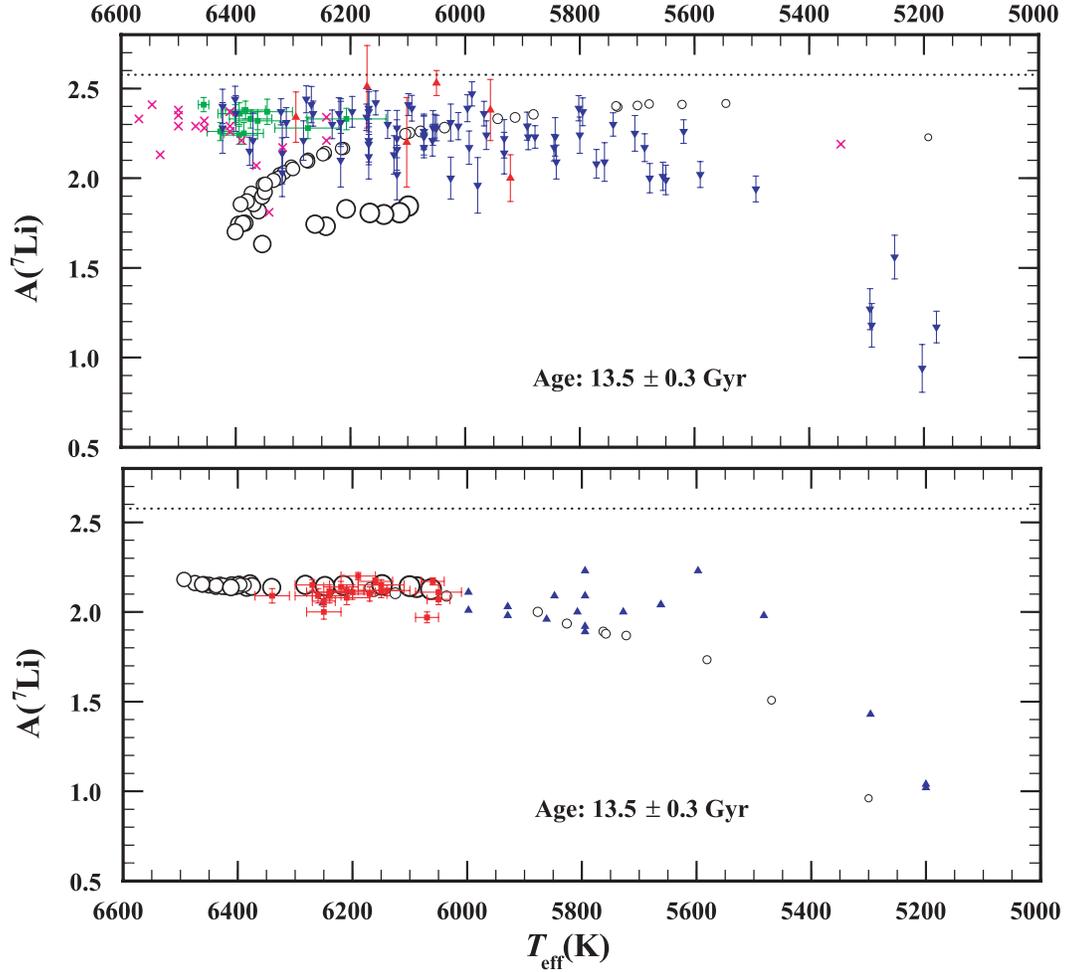}}
\caption{Predicted Li abundance in stars without turbulence (upper part) and with the T6.25 turbulence (lower part). 
For the calculations, the initial value of A($^7$Li) is 2.58 
and open black circles are used.  The size of the circles is a function of the radius of the stars in order
to indicate roughly their evolutionary stage. There are  50 simulated stars in each panel and 90 \% of them are between 13 and 14 Gyr.
Further description is found in the text.  Also shown
are observations in metal poor halo stars in the upper part by \citet{Thorburn94} (blue triangles), 
\citet{BonifacioMoSietal2003} (magenta crosses), \citet{BonifacioPaSpetal2002} (green squares), \citet{Bonifacio2002} (red triangles)
  and in the lower 
part by \citet{SpiteMaSp84} (blue filled triangles) and by \citet{RyanNoBe99} (red filled squares with error bars).
}
\label{fig:casino}
\end{figure}

We have learned a number of properties of the evolutionary models of Pop II stars that make this constancy less surprising 
than it appeared when first noticed. What is needed to reproduce the Spite plateau may be inferred from Figure \ref{fig:prof_li}.
A small amount of turbulence below the surface convection zone (the T6.0 or perhaps the T6.09 model) is all that is required
to produce a sufficiently flat Li abundance plateau.  Whether turbulent transport needs to extend deeper in, to the Li burning region,
is currently uncertain.  This extension would cause the destruction of the Li peak seen at $\log T = 6.25$ in Figure \ref{fig:prof_li}.
That turbulence does not directly modify gravitational settling from the convection zone.
In the absence of turbulence, the value of the Li concentration in that peak is determined by atomic diffusion but 
is drastically reduced by any turbulence 
between $\log T = 6.25$ and 6.40.  A reduction of this peak is needed if the Spite plateau is more than a factor of 1.6 to 2 below 
the cosmological Li abundance given by equation (\ref{eq:wmapli}) so that gravitational settling from the convection zone may 
proceed without being excessively reduced by a backward turbulent diffusion transport (see Fig. \ref {fig:vdif}).
If the high \teff{} scale is the right one, the Li reduction factor from the WMAP value can probably be 
explained without a reduction of this peak.

If, on the other hand, the low \teff{} scale is the right one,  turbulent transport must approach the T6.25 model.  It then plays
a role in destroying the Li peak mentioned above.  The T6.25 turbulent transport model is  surprisingly similar 
to the turbulent transport 
which is needed to reproduce the Li abundance reduction in the Sun (see \S \ref{sec:ZC}).
The  low temperature 
scale however leads to some difficulties with cluster isochrones \citep{VandenBergRiMietal2002}.

It is often assumed that the Pop II stars with a Li abundance a factor of 2 or more below the Spite plateau and a $\teff{}> 6000$ K
 (see for instance \citealt{BonifacioMoSietal2003} and \citealt{RyanBeKaetal2001})
are caused by a different process and should not be included in the search for an explanation of this plateau.
If one looks at the upper part of Figure~\ref{fig:casino} one notes that, in the absence of turbulence, 
post turnoff stars could have a factor of 2 to 3 lower surface abundance than pre turnoff stars.  Such objects appear 
much less numerous than the number of post turnoff stars expected in this simulation where all stars are assumed
to have no turbulence below the convection zone.  It is however tempting to make the
same suggestion as we made (see \S 5 of \citealt{RichardMiRietal2002}) to explain the \citet{KingStBoetal98} 
observations of Fe abundance variations in some M\,92 turnoff stars.  Just as in Pop I, some 20~\% of A stars have 
lower turbulence than most and develop abundance anomalies by atomic diffusion processes \citep{RicherMiTu2000}, 
similarly a fraction (say 20~\%) of Pop II turnoff stars could have lower turbulence and have anomalously small Li.  This may
be tested by a confirmation of the abundance variations in post turnoff stars of M\,92.  While \citet{KingStBoetal98} 
suggest the presence of such anomalies, higher signal to noise observations appear needed to confirm this result especially since
such anomalies are not seen in higher metallicity clusters, where however they would not be expected to form by 
atomic diffusion according to Figure 10 of \citet{RichardMiRi2002}. 

The surface abundance of $^6\Li$ in the hotter stars of the plateau offers another test of the role of transport processes. 
While consistent with the transport being due to atomic diffusion, as described in \S \ref{sec:settling},
 the presence of $^6\Li$ appears to rule out 
the T6.25 model, that is  the most turbulent transport model consistent with WMAP and the Spite plateau (see \S \ref{sec:discussion}).  
It is then a very important additional constraint 
on turbulent transport and
it would be important to confirm and or extend  the few positive $^6\Li$ detections mentioned in \S \ref{sec:observations}.

Given that the WMAP implications for Li cosmological abundance and the Li Spite plateau can be naturally 
explained by gravitational settling in presence of weak turbulence, there appears little need for exotic physics 
as suggested by 
\citet{IchikawaKaTa2004} and \citet{CocVaDeetal2004} for instance.
Instead, there is a need for a better understanding of turbulent transport in the radiative zones of stars.
This requires simulations  from first principles such as are being attempted for instance 
 by \citet{TalonViMietal2003} and \citet{TheadoVa2003}.

\acknowledgments
The authors thank D. VandenBerg and A. Babul for proding them into writing this paper.
We thank an anonymous referee for useful remarks which led to our adding \S 3.3.1.
This research was partially supported at  the Universit\'e de Montr\'eal 
by NSERC. We thank the R\'eseau Qu\'eb\'ecois de Calcul de Haute
Performance (RQCHP)
for providing us with the computational resources required for this
work.




%\bibliography{astrojabb,michaud}
%\bibliographystyle{apj}
%\input{ms.bbl}

%% the file, uncomment the \plotone, \plottwo, and
%% If you need a layout that cannot be achieved with \plotone or
%% \plottwo, you can invoke the graphicx package directly with the
%% \includegraphics command or use \plotfiddle. For more information,
%% please see the tutorial on "Using Electronic Art with AASTeX" in the
%% documentation section at the AASTeX Web site,
%% http://www.journals.uchicago.edu/AAS/AASTeX.
%%
%% The examples below also include sample markup for submission of
%% supplemental electronic materials. As always, be sure to check
%% the instructions to authors for the journal you are submitting to
%% for specific submissions guidelines as they vary from

%% This example uses \plotone to include an EPS file scaled to
%% 80% of its natural size with \epsscale. Its caption
%% has been written to indicate that additional figure parts will be
%% available in the electronic journal.

\end{document}